\documentclass[aps,showpacs, twocolumn]{revtex4}
\usepackage{amsmath,latexsym} 
\begin{document}

\preprint{UCMFTEOR-01- , Missouri-Columbia-01-}
\title{\bf A DECOUPLED SYSTEM OF HYPERBOLIC EQUATIONS FOR LINEARIZED 
COSMOLOGICAL PERTURBATIONS}

\author{J. Ram\'{\i}rez \footnote{Email address: mittel@foton.fis.ucm.es} }
\affiliation{Departamento de Fisica Teorica I, Universidad Complutense de Madrid,
28040 Madrid, Spain.}

\author{S. Kopeikin \footnote{Email address: kopeikins@missouri.edu} }
\affiliation{Department of Physics and Astronomy, 
University of Missouri-Columbia, Columbia, MO 65211, USA.}

\date{\today}

\begin{abstract}

A decoupled system of hyperbolic partial differential equations for 
linear perturbations around any spatially flat FRW universe is obtained 
for a wide class of perturbations. The considered 
perturbing energy momentum-tensors
can be expressed as the sum of the perturbation of a
minimally coupled scalar field plus an arbitrary (weak) energy-momentum tensor
which is covariantly conserved with respect to the background. The key
ingredient in obtaining the decoupling of the equations is the introduction
of a new covariant gauge which plays a similar role as harmonic gauge
does for perturbations around Minkowski space-time. The case of universes
satysfying a linear equation of state is discussed  in particular, and
closed analytic expressions for the retarded Green's functions 
solving the de Sitter, dust and radiation dominated cases are given.


\end{abstract}

\maketitle

{\bf 1. Introduction}
\vskip 0.15cm
The theory of linearized perturbations around cosmological backgrounds is 
a cornerstone in the understanding of structure formation in the universe,
because it provides the connection between what we see today, and the initial 
conditions in the early universe \cite{Bertschinger}. This theory has a
long history, beginning (in the context of general relativity)
with the pioneering work of E. Lifshitz \cite{Lika}. An important step was
the  introduction by  J. Bardeen of gauge invariant potentials to
describe the metric perturbations \cite{Bardeen}. This is the so called
standard treatment of cosmological perturbations, and many authors
have elaborated  this approach  and  used it to deal with the analysis of 
observations \cite{Mukha}. The standard treatment is based
on the decomposition of the metric perturbations in scalar, vector and 
tensor parts according to their behaviour under the rotation group
of isometries of the FRW background \cite{isom}. In the standard approach
these three kinds of 
perturbations are treated independently, and most attention has been 
devoted to the discussion of scalar perturbations in the particular case 
in which the two gauge invariant scalar potentials coincide. 
This particular case is well motivated because it arises naturally when 
dealing with hydrodynamic or scalar field perturbations.  In these 
cases, the space-space components of the perturbing energy momentum-tensor 
take the form $\delta T_i^j \propto S\, \delta_i^j$, where $S$ is a scalar, 
and in turn this implies the equality $\Phi = \Psi$, between the two gauge 
invariant Bardeen potentials for the scalar sector. 
An important technical consequence of this 
equality is that one has to deal only with a second order 
differential equation
for $\Phi$; in contrast with the general case of scalar perturbations, 
in which the standard formalism yields two coupled equations for  
$\Phi$ and $\Psi$ \cite{Mukha}. 
 
In this letter we generalize  previous results that we first obtained  for
a dust filled universe \cite{dust}, and  present a new approach to 
cosmological perturbations, 
which allows a unified treatment of all types of perturbations: scalar, vector
and tensor.  The method follows a very similar
path to that used for computing the  gravitational radiation 
emitted by astrophysical sources in Minkowski space-time, using harmonic
gauge \cite{Weinberg}. As a result of our approach, we obtain a 
decoupled system of hyperbolic partial differential equations for all 
components of the metric perturbation $h_{\mu \nu}$.  
This decoupling of equations  holds for
a very wide class of perturbations, which we define below, including the
presence of seeds like cosmic strings or primordial black holes
\cite{Durrer}. In addition the equations for perturbations are formulated in
a covariant way, and in our opinion they are best fitted to study linearized
quantum gravity on spatially flat cosmological backgrounds. 

\vspace{0.75cm}

{\bf 2. The background model}
\vskip 0.15cm 
We choose a spatially flat FRW universe as background. This simplifies
matters a little, at not a very high cost concerning  physics,  
because the recent 
analysis of observations of supernova type Ia \cite{supernova}  and 
of acoustic peaks in the distribution of CMB temperature versus angular 
momentum \cite{Boomerang}, favour a flat universe. In addition, inflationary 
models of the very early universe also favour flat universes \cite{Linde}. The
metric of such background model written in conformal time $\eta$ and spatial
cartesian coordinates $\vec{x}$, reads \cite{conventions} 

\begin{equation}
ds^2 = \gamma_{\mu \nu} \,  dx^{\mu}dx^{\nu} = a(\eta)^2 \,  
(- d\eta^2 + d\vec{x}^2) ~~,
\label{backmet}
\end{equation}

\noindent
where the scale factor $a(\eta)$ is an arbitrary function
of time. Thus, the equations for perturbations that we are going to derive
will apply to any flat FRW 
background, and in particular to multifluid universes, like  
our own universe.
To describe the matter producing this background, we shall use
a scalar field $\phi$ minimally coupled to gravity, i.e. we shall
consider the action

\begin{eqnarray}
 S\left[g_{\alpha \beta}, \phi \right] &  =  &  
- \frac{1}{16 \pi {\mathcal{G}} } \int d^4 x \, \sqrt{-g} \,  R \nonumber \\  
& &  - \int d^4 x \, \sqrt{-g} \left( \frac{1}{2}\, g^{\alpha \beta}\, 
\phi,_{\alpha} \, \phi,_{\beta}\,  + \, V(\phi) \right) ~~.
\label{action}
\end{eqnarray}

The full metric $g_{\alpha \beta}$ and the scalar field $\phi$ are to be 
developed up to first order in perturbations, in the form 
 $g_{\alpha \beta} = \gamma_{\alpha \beta} + h_{\alpha \beta}$, and
$\phi = \bar{\phi} + \delta \phi$. The  zero order parts
describe the background. It is important to emphasize that this description
of the background, by means of the scalar field $\phi$, does not place 
any restriction on the scale factor $a(\eta)$.  Any desired function
$a(\eta)$ can be obtained
by an appropriate choice of the potential $V(\phi)$.
In this sense, the scalar field $\phi$ is not necessarily to be thought 
of as a fundamental field filling the universe, but as a convenient way of 
parametrizing the matter content of the universe at cosmological scales;
as for example cold matter plus a cosmological constant, or whatever
model future observations will support. 

Using this scalar field parametrization of the matter, the 
zeroth order energy-momentum tensor can be written in the
perfect fluid covariant form

\begin{equation}
\bar{T}_{\mu}^{\hspace{0.2cm} \nu} = \left[\frac{1}{2} \, \bar{u}^{\alpha} 
\bar{u}^{\beta}
\bar{\phi},_{\alpha} \bar{\phi},_{\beta} \,  - \, V(\bar{\phi}) 
\right] \delta_{\mu}^{\nu} \,  + \, 
\bar{u}^{\alpha} \bar{u}^{\beta} \bar{\phi},_{\alpha} \bar{\phi},_{\beta}  
\, \bar{u}_{\mu} \bar{u}^{\nu} ~~,
\label{Tzero}
\end{equation}

\noindent
where $\bar{u}^{\nu}$ is the background velocity field of the fluid. 
The values of the background pressure and density can
be derived from (\ref{Tzero}), and the covariant conservation of 
$\bar{T}_{\mu}^{\nu}$, with respect to the background metric,
yields the  zeroth order equation of motion for the field $\bar{\phi}$ 

\begin{equation}
\bar{\phi}_{|\alpha}^{\hphantom{|\alpha} |\alpha} = 
\frac{\partial V\left(\bar{\phi}\right)}{\partial \bar{\phi}} ~~,
\label{eqfi}
\end{equation}
\noindent
where the vertical bar $|$ denotes the covariant derivative with respect
to the background metric.

In coordinates $(\eta, \vec{x})$, and using a dot for derivatives with 
respect 
to conformal time $\eta$,  the background values of pressure,
density, and velocity take the form:

\begin{equation}
\bar{p} = \frac{1}{2}\,  a^{-2}\,  {\Dot{\bar{\phi}}}^{\hspace{0.1cm} 2} - 
V\left(\bar{\phi}\right) ~~,
\label{preszero}
\end{equation}

\vspace{-0.4cm}

\begin{equation}
\bar{\rho} = \frac{1}{2}\,  a^{-2}\,  {\Dot{\bar{\phi}}}^{\hspace{0.1cm} 2}
 + V\left(\bar{\phi}\right) ~~,
\label{denszero}
\end{equation}

\noindent
and
\vspace{-0.4cm}

\begin{equation}
\bar{u}^{\nu}   = a^{-1} \,  \delta^{\nu}_0~~.
\label{velzero}
\end{equation}

\noindent
Writting $a = \exp \Omega$, the Hubble parameter is given by

\begin{equation}
H = \frac{\Dot{a}}{a^2} = \Dot{\Omega} \,  \exp (-\Omega) ~~,
\label{Hubble}
\end{equation}

\noindent
the Einstein equations for the background are

\begin{equation}
\Ddot{\Omega} - {\Dot{\Omega}}^2 = - 4 \pi {\mathcal{G}} \,
{\Dot{\bar{\phi}}}^{\hspace{0.1cm} 2}  ~~, 
\label{Einseqback1}
\end{equation}

\begin{equation}
\Ddot{\Omega} \,  + \, \frac{1}{2} \, {\Dot{\Omega}}^2 = - 4 \pi {\mathcal{G}}
\, \left[ {\frac{1}{2}\, \Dot{\bar{\phi}}}^{\hspace{0.1cm} 2}  - 
a^2 V\left(\bar{\phi}\right) 
\right] ~~, 
\label{Einseqback2}
\end{equation}

\noindent
and the integrability condition (\ref{eqfi}) reads

\begin{equation}
\Ddot{\bar{\phi}} \, + \, 2\, \Dot{\Omega} \, \Dot{\bar{\phi}} \, 
+  \, a^2 \,   \frac{\partial V\left(\bar{\phi}\right)}{\partial \bar{\phi}}
= 0  ~~.
\label{eqfieta}
\end{equation}

For example, in the particular case of a flat FRW background
with a linear equation of state $p = \alpha \, \rho$, 
the corresponding potential for the scalar field is

\begin{equation} 
V(\phi) = \frac{3 (1-\alpha)\, H_0^2}{16 \pi {\mathcal{G}}}
\, \exp \left( -2 \sqrt{6 (1 + \alpha ) \pi {\mathcal{G}}} \, (\phi - \phi_0)
\right) ~~,
\label{potential}
\end{equation}

\noindent
and the solutions of the  Einstein equations for the Hubble 
parameter, the background scalar field $\bar{\phi}$, and
the conformal Hubble parameter $\Dot{\Omega}$, are  

\begin{equation} 
H(\eta) = 
H_0 \left( \frac{\eta_0}{\eta} \right)^{
\displaystyle \frac{3 + 3 \alpha}{1 + 3 \alpha}}
~~,
\label{easyHubble}
\end{equation}

\begin{equation}
\bar{\phi} = \bar{\phi}_0  \, + \,
\frac{1}{1+3\alpha} \sqrt{\frac{3 + 3 \alpha}{2 \pi {\mathcal{G}}}} \,
\ln \frac{\eta}{\eta_0}~~, 
\label{easyfiback}
\end{equation}

\noindent
and
\vspace{-0.4cm}

\begin{equation}
\Dot{\Omega} = \frac{2}{(1 + 3 \alpha ) \, \eta}
\label{easyconfHubble}~~,
\end{equation}

\noindent 
the 0 label refers, as usual, to the present epoch values of the
corresponding quantities, and the evolution of the density follows the 
critical value law $\rho = 3 H^2/8 \pi {\mathcal{G}}$.

\vspace{0.75cm}

{\bf 3. Structure of perturbations and gauge invariance}
\vskip 0.15cm
Now we are going to specify the class of perturbations that we will
allow on the background. The equations for perturbations come out, as
usual, by developing the Einstein tensor up to first order
in perturbations 
$ G_{\mu}^{\nu} = \bar{G}_{\mu}^{\nu} + \delta G_{\mu}^{\nu}$,
and splitting the Einstein equations into a zeroth order part

\vspace{-0.2cm}
\begin{equation}
\bar{G}_{\mu}^{\hphantom{\mu}\nu} = - 8  \pi {\mathcal{G}} \, 
\bar{T}_{\mu}^{\hphantom{\mu}\nu} ~~,
\label{Einseqcero}
\end{equation}

\noindent
which describes how the background is created by  $\bar{T}_{\mu}^{\nu}$,
and a first order part  

\begin{equation}
\delta G_{\mu}^{\hphantom{\mu} \nu} = - 8  \pi {\mathcal{G}} \, 
\delta T_{\mu}^{\hphantom{\mu}  \nu} ~~,
\label{Einsequno}
\end{equation}

\noindent
which gives the equation governing perturbations.

Expansion of the Einstein tensor up to first order in peturbations yields
the covariant expression

\begin{eqnarray}
2 \,  \delta G_{\mu}^{ \hphantom{\mu} \nu} & = & \psi_{\mu  \hphantom{\nu}  
|\lambda \hphantom{\lambda}}^{\hphantom{\mu} \nu \hphantom{\lambda} |\lambda} 
\,  - \,  \psi_{\mu \hphantom{\lambda} |\lambda}^{\hphantom{\mu} \lambda
 \hphantom{\lambda} |\nu} 
\, - \, \psi_{\hphantom{\nu} \lambda \hphantom{| \lambda} |\mu}^{\nu 
\hphantom{\lambda} |\lambda}
\, + \, \delta_{\mu}^{\nu} \, 
\psi_{\alpha \hphantom{\beta} |\beta}^{\hphantom{| \alpha} \beta 
\hphantom{\beta}  |\alpha}
\nonumber \\
& & + \,  2\, \bar{R}^{\nu}_{\hphantom{\nu} \alpha \beta \mu}  
\, \psi^{\alpha \beta}
\, + \, \bar{R}_{\alpha}^{\hphantom{\alpha} \nu} \, 
\psi_{\mu}^{\hphantom{\mu} \alpha}
\, - \, \bar{R}_{\mu}^{\hphantom{\alpha} \alpha} \, 
\psi_{\alpha}^{\hphantom{\alpha} \nu}
\nonumber \\
& & + \, \bar{R}_{\mu}^{\hphantom{\mu} \nu} \, \psi \,  + \, 
\delta_{\mu}^{\nu}
\left[\bar{R}_{\alpha \beta} \, \psi^{\alpha \beta} 
\, - \, \frac{1}{2} \, \bar{R} \, \psi \right] ~~, 
\label{Einsequnocov}
\end{eqnarray}

\noindent
where $\psi_{\mu}^{\, \nu} = h_{\mu}^{\, \nu} - 
\frac{1}{2}  \delta_{\mu}^{\nu} \,  h$ is the so called trace reversed
graviton field.
On the other hand the structure of the perturbing energy-momentum tensor 
$ \delta{T}_{\mu}^{\hphantom{\mu} \nu}$ 
is constrained by the Bianchi identity. Expanding the full Bianchi identity
to first order in perturbations, and taking into account the covariant
conservation of $\bar{T}_{\mu}^{\hphantom{\mu} \nu}$ with respect to the 
background metric, one obtains the constraint

\begin{equation}
\delta T_{\mu \hphantom{\nu} |\nu}^{\hphantom{\mu}\nu} \, + \, 
\delta \Gamma_{\nu \alpha}^{\nu}\, \bar{T}_{\mu}^{\hphantom{\mu} \alpha} - 
\delta \Gamma_{\nu \mu}^{\alpha} \, \bar{T}_{\alpha}^{\hphantom{\alpha} \nu}  
= 0 ~~,
\label{Bconstraint}
\end{equation}

\noindent
where $\delta \Gamma_{\mu \nu}^{\lambda}$ is the first order part of the full
metric connection.  The inhomogeneous equation (\ref{Bconstraint}) must be 
fulfilled by any perturbing energy momentum tensor. Thus, any perturbing 
$\delta T_{\mu}^{\nu}$ can be decomposed in the form

\begin{equation}
\delta  T_{\mu}^{\nu} =  \delta T_{\mu}^{(I) \nu} + \,
\delta T_{\mu}^{(F) \nu}  ~~. 
\label{Tdecomp}
\end{equation}

\noindent
where $\delta T_{\mu}^{(I) \nu}$ is a particular solution of the constraint 
equation (\ref{Bconstraint}), and $\delta T_{\mu}^{(F) \nu}$ is any solution
of the homogeneous equation 

\vspace{-0.2cm}
\begin{equation}
\delta T^{(F)\nu}_{\mu |\nu} = 0 ~~. 
\label{covcons}
\end{equation}

In addition, we will assume that $\delta T_{\mu}^{(F) \nu}$ does not 
functionally depend on the metric perturbation $h_{\mu \nu}$.
We shall call $\delta T_{\mu}^{(I) \nu}$ the {\em intrinsic} perturbation,
and  $\delta T_{\mu}^{(F) \nu}$ the {\em free} perturbation \cite{names}.

From the physical point of view, the 
intrinsic perturbation $\delta T_{\mu}^{(I) \nu}$ corresponds to the
irregularities in the matter that creates the background, i.e. to deviations
of homogeneity and isotropy in the matter which produces the background
metric. On the other hand, what we call the free perturbation corresponds
to additional matter (more or less exotic) that can exist in the
universe in addition to the main component, and which moves as test matter
on the background, i. e. it fulfills equation (\ref{covcons}). This 
extra matter could be, for instance, topological defects as cosmic
strings or domain walls produced by phase transitions in the early
universe \cite{strings}, or any other seed perturbation as for example
primordial black holes. But also, ordinary baryonic matter, 
amounting only to 
two or three percent of the total energy density in the present universe, 
can be considered as a free perturbation too. In the last case the
intrinsic perturbation $\delta T_{\mu}^{(I) \nu}$  would correspond entirely 
to nonbaryonic dark matter or energy.
  
Now the expression for $\delta T_{\mu}^{(I) \nu}$ is obtained from the action
(\ref{action}), and in turn, it can be decomposed as 

\begin{equation}
\delta  T_{\mu}^{(I) \nu} = \delta  T_{\mu}^{(h) \nu} + \, 
\delta  T_{\mu}^{(\phi) \nu} ~~,
\label{decompTI}
\end{equation}

\noindent
where $\delta  T_{\mu}^{(h) \nu}$ and $\delta  T_{\mu}^{(\phi) \nu}$
are the pieces linear in the metric perturbation $h_{\mu \nu}$, and in
the scalar field perturbation $\delta \phi$ respectively.
The covariant form of these pieces is

\begin{equation}
\delta  T_{\mu}^{(h) \nu} = (\bar{p} + \bar{\rho})
\left[ \frac{1}{2} \, \bar{u}_{\alpha} \bar{u}_{\beta} \, h^{\alpha \beta}
 \, \delta_{\mu}^{\nu} - \bar{u}_{\mu} \bar{u}_{\alpha} \,
h^{\alpha \nu} \right]  ~~,
\label{Th}
\end{equation}

and
\vspace{-0.4cm}

\begin{eqnarray} 
\delta  T_{\mu}^{(\phi) \nu}  &  = &  \gamma^{\nu \alpha}
\left[ \bar{\phi},_{\mu} \,  \delta \phi,_{\alpha} + \,
  \bar{\phi},_{\alpha} \,  \delta \phi,_{\mu} \right] \nonumber \\ 
& &  - \, \delta_{\mu}^{\nu} \left[ \gamma^{\alpha \beta} \, 
  \bar{\phi},_{\alpha} \,  \delta \phi,_{\beta} \,  - \,
\frac{\partial V\left(\bar{\phi}\right)}{\partial \bar{\phi}} \, 
\delta \phi \right]   ~~.
\label{Tfi}
\end{eqnarray}

Let us discuss now how the perturbations behave under gauge transformations
induced by  infinitesimal coordinate transformations. Given an infinitesimal
coordinate transformation $ x^{\mu}  \longrightarrow 
 x^{\prime \mu}  =  x^{\mu} - \xi^{\mu} (x)$, the corresponding gauge 
transformations of the metric and scalar field perturbations are 

\begin{equation} 
h_{\mu \nu} \longrightarrow h^{\prime}_{\mu \nu} = 
h_{\mu \nu} \,  + \,  \xi_{\mu |\nu} \,  + \, \xi_{\nu |\mu}  ~~,
\label{gaugeh}
\end{equation}

\noindent
and
\vspace{-0.4cm}

\begin{equation}
\delta \phi   \longrightarrow \delta \phi^{\prime} = 
\delta \phi \, + \, \xi^{\mu} \bar{\phi},_{\mu} ~~.
\label{gaugefi}
\end{equation}

\noindent
Then, the Einstein tensor perturbation and the intrinsic energy-momentum
perturbation transform accordingly as

\begin{equation}
\delta G_{\mu}^{\hphantom{\mu}\nu} \longrightarrow 
\delta G_{\mu}^{\prime \hphantom{\mu} \nu} =
\delta G_{\mu}^{\hphantom{\mu} \nu}  + \, {\mathcal{L}}_\xi \,
{\bar{G}}_{\mu}^{\hphantom{\mu} \nu} ~~. 
\label{gaugeG}
\end{equation}

\noindent
and
\vspace{-0.4cm}

\begin{equation}
\delta T_{\mu}^{(I) \nu} \longrightarrow 
\delta T_{\mu}^{(I) \prime  \nu} =
\delta T_{\mu}^{(I) \nu}   + \,
{\mathcal{L}}_\xi \,  {\bar{T}}_{\mu}^{\hphantom{\mu} \nu} ~~, 
\label{gaugeT}
\end{equation}

\noindent
where $\mathcal{L}_\xi$ is the Lie derivative with respect to the
infinitesimal vector field $\xi^{\mu}$. We have assumed, as usual, that 
the zeroth order parts of all tensors remain invariant in a fixed coordinate 
system on the background space-time manifold 
(for example  $(\eta , \vec{x})$, but not necessarily this one), and 
that the effect of infinitesimal coordinate transformations is fully charged 
to the perturbations of tensors.  Also notice that the transformation laws 
(\ref{gaugeG}) and (\ref{gaugeT}), can be directly checked by
replacing the transformation laws (\ref{gaugeh}) and (\ref{gaugefi})
into the functional expressions (\ref{Einsequnocov}), (\ref{Th})  and 
(\ref{Tfi}).

It is important to observe that the free perturbation 
$ \delta T_{\mu}^{(F) \nu}$ is gauge invariant. This is compulsory because 
the total perturbation $ \delta T_{\mu}^{\hphantom{\mu} \nu}$ 
has the same transformation law (\ref{gaugeT}) as the intrinsic perturbation 
$ \delta T_{\mu}^{(I) \nu}$. This happens due to the fact that 
the gauge transformation of $ \delta T_{\mu}^{(F) \nu}$ is  of 
second order in perturbations, since this energy-momentum tensor 
has no zeroth order counterpart. Moreover, the zeroth order
Einstein equations (\ref{Einseqcero}) imply that the combination

\begin{equation}
\hat{\delta G}_{\mu}^{\hphantom{\mu} \nu} \equiv \delta G_{\mu}^{\hphantom \nu}
+  \, 8  \pi {\mathcal{G}} \,  \delta T_{\mu}^{(I) \nu} ~~,
\label{gaugeinvariantG}
\end{equation}

\noindent
is gauge invariant. Therefore, by sending the intrinsic perturbation to
the l.h.s. of (\ref{Einsequno}), we obtain the equations

\begin{equation}
\hat{ \delta G}_{\mu}^{\hphantom{\mu} \nu} = - \,  8  \pi {\mathcal{G}} \, 
 \delta T_{\mu}^{(F) \nu} 
\label{pertcoveq}
\end{equation}

\noindent
in which both sides are gauge invariant. Thus equations (\ref{pertcoveq})
for cosmological perturbations are covariant with respect to finite general 
coordinate transformations, and gauge invariant with respect to
infinitesimal coordinate changes around a fixed coordinate system
on the background space-time.
 
Also notice that both sides of (\ref{pertcoveq}), are covariantly conserved
with respect to the background metric.

\vspace{0.75cm}
      
{\bf 4.  Choosing the gauge and decoupling of the equations}
\vskip 0.15cm
Now we are finally going to give the covariant and gauge invariant equations
(\ref{pertcoveq}) a nice definite form. This entails a combination of two
things: choosing a convenient coordinate system on the background, and fixing 
the gauge.
Although all gauges contain the same physics, most of them contain it
in a complicated way, meaning that the independent functions which 
solve the equations that remain after the gauge fixing, have spurious
complications related to the gauge, but not related to real complications 
in the physics. The situation here is similar to that of   
the simple system of gauge invariant Maxwell equations
for the photon field $\Box A_{\mu} - \partial_{\mu} (\partial_{\nu} A^{\nu})
= J_{\mu}$. In Lorentz gauge ($\partial_{\nu} A^{\nu} = 0$), one has nice 
solutions for these equations, but one could complicate the functional form of 
the solutions by choosing an inappropriate (but possible) 
gauge fixing condition.

In the case of the equations for cosmological perturbations
(\ref{pertcoveq}), the appropriate gauge simplifying the equations
is not so obvious as in the case of electrodynamics. 
The strategy that we follow is inspired by the way in which
harmonic gauge works to simplify the equations for perturbations around
Minkowski space-time. First, we set the covariant gauge fixing condition 

\vspace{-0.2cm}
\begin{equation}
\psi_{\mu \hphantom{\nu} |\nu}^{\hphantom{\mu} \nu} = B_{\mu} ~~, 
\label{covgaugecond}
\end{equation}

\noindent
where the field $B_{\mu}$ is yet unspecified. Then, if $B_{\mu}$  does not
depend on the derivatives of the metric perturbation 
$\psi_{\mu \hphantom{\nu} |\nu}$, 
this condition eliminates all terms in second derivatives coming from
the second, third and fourth terms in the expression (\ref{Einsequnocov}) for 
$\delta G_{\mu}^{\hphantom{\mu} \nu}$.  
Thus, only the flat d'Alembertian remains as second order differential 
operator in the left hand side of (\ref{pertcoveq}). In addition, working in
coordinates $(\eta, \vec{x})$, we
fix the $B^{\mu}$ field to simplify the equations (\ref{pertcoveq}) as
much as possible. After the long but straightforward exercise of 
writing
$\hat{\delta G}_{\mu}^{\hphantom{\mu} \nu}$ given by the covariant expressions 
(\ref{Einsequnocov}), (\ref{Th}), (\ref{Tfi}) and (\ref{gaugeinvariantG}) 
in coordinates  $(\eta, \vec{x})$, the analysis of the resulting expression  
shows that a great simplification is achieved by the choice 

\vspace{-0.2cm}
\begin{equation}
B_{\mu} =  - \,  2 H \, \bar{u}_{\nu} \, \psi_{\mu}^{\hphantom{\mu} \nu} + \, 
16 \pi {\mathcal{G}} \, \bar{\phi},_{\mu} \, \delta \phi ~~, 
\label{Bmu}
\end{equation}

\noindent
which has also been written in covariant form. Then, using the choice
(\ref{Bmu}) to specify the gauge (\ref{covgaugecond}), the cosmological 
perturbations equations (\ref{pertcoveq})
in this gauge, and in coordinates $(\eta, \vec{x})$,
take the simple form 
\begin{widetext}
\begin{equation}
\Box \psi_{\mu}^{\hphantom{\mu} \nu} - \, 2\, \Dot{\Omega} \, \partial_{\eta}
\psi_{\mu}^{\hphantom{\mu} \nu} + \,  2 \Ddot{\Omega} \, 
\left[ \delta_{\mu}^0 \,  \delta_0^{\nu} \,  \psi - \, 
\delta_{\mu}^0 \, \psi_0^{\hphantom{0} \nu}
- \, \delta_0^{\nu} \, \psi_{\mu}^{\hphantom{\mu} 0} \right] \,
+ 32 \pi {\mathcal{G}} \, \delta_{\mu}^0 \,  \delta_0^{\nu} \, 
\Ddot{\bar{\phi}} \, \delta \phi = \, - 16  \pi {\mathcal{G}} \, a^2 \,  
\delta T_{\mu}^{(F) \nu} ~~, 
\label{pertetaeq}
\end{equation}
\end{widetext}

\noindent
where $\Box = - \partial^2 / \partial \eta^2 + \, {\vec{\nabla}}^2$ is 
the Minkowski d'Alembertian.

In addition the covariant conservation of $\delta T_{\mu}^{(F) \nu}$,
when applied to equations (\ref{pertetaeq})   
yields an equation for the 
perturbation $\delta \phi$, which can also be obtained by perturbing the
full equation of motion for the field $\phi$.  This last equation can be 
skipped since it is implicit in the equations (\ref{pertetaeq}), as can
be checked by direct computation and taking into account the gauge 
fixing condition. Instead the scalar field perturbation $ \delta \phi$
can be expressed in terms of the metric perturbations through the
gauge fixing condition. In coordinates  $(\eta, \vec{x})$, we obtain

\begin{equation}
\delta \phi = \frac{1}{16 \pi {\mathcal{G}} \, \Dot{\bar{\phi}}}
\left[ \partial_{\nu} \psi_0^{\hphantom{0} \nu} + \, 
\Dot{\Omega}\,  (2 \, \psi_0^0 - \,  \psi) \right]
\label{gaugecondeta}
\end{equation}

Finally, replacing (\ref{gaugecondeta}) in (\ref{pertetaeq}), these ten 
equations can be recast in the form 

\begin{equation}
\left[ \Box \, + \, {\Dot{\Omega}}^2 + \, \Ddot{\Omega} \right]
 \left( a \, \psi_i^{\hphantom{i} j} \right) = - \,  16  \pi {\mathcal{G}}
\, a^3 \,  \delta T_{i}^{(F) j} ~~, 
\label{cojoeqss}
\end{equation}

\vspace{-0.4cm}
\begin{equation}
\left[ \Box \, + \, {\Dot{\Omega}}^2 - \, \Ddot{\Omega} \right]
 \left( a \, \psi_0^{\hphantom{0} j} \right) = - \,   16  \pi {\mathcal{G}}
\, a^3 \,  \delta T_{0}^{(F) j} ~~,
\label{cojoeqts}
\end{equation}

\noindent
and
\begin{widetext}
\begin{equation}
\left[ \Box \, + \, {\Dot{\Omega}}^2 - \, \Ddot{\Omega} \, 
+ \, \left( \frac{\Ddot{\bar{\phi}}}{\Dot{\bar{\phi}}} \right)^2  \,
- \, \partial_{\eta} \left( \frac{\Ddot{\bar{\phi}}}{\Dot{\bar{\phi}}} \right) 
\right] \left( \frac{a}{\Dot{\bar{\phi}}} \, 
(2 \psi_0^{\hphantom{0} 0} - \, \psi) \right) = \,
- 16  \pi {\mathcal{G}} \, \frac{a^3}{\Dot{\bar{\phi}}} \,
(2 \,  \delta T_{0}^{(F) 0} - \delta T^{(F)} ) - \, 
2 \, \frac{\Ddot{\bar{\phi}}}{{\Dot{\bar{\phi}}}^2} \, a \, 
(\partial_{\eta} \psi_i^{\hphantom{i} i} 
+  \partial_i  \psi_{0}^{\hphantom{0} i}) ~~.
\label{cojoeqtt}
\end{equation}
\end{widetext}

Equations (\ref{cojoeqss}),  (\ref{cojoeqts}) and (\ref{cojoeqtt}) are a 
decoupled system of hyperbolic partial 
differential equations for the ten components of the metric perturbation.
They can be exactly integrated and their solutions expressed as retarded
potentials using the retarded 
Green's functions for the three differential operators appearing on the 
left hand sides.
Equations  (\ref{cojoeqss}) and (\ref{cojoeqts}), for the $\psi_i^j$ and 
$\psi_0^i$ components, 
must be solved first  because some combinations of the spatial components 
of the metric perturbation enter as a source in equation (\ref{cojoeqtt}).

The decoupled system of hyperbolic equations (\ref{cojoeqss}),  (\ref{cojoeqts}) 
and (\ref{cojoeqtt}) is the central result of this paper. They give a unified
description of all three types of metric perturbations: scalar, vector and 
tensor, and they have a high degree of generality since they are valid for
any spatially flat FRW background and any perturbation that can be 
decomposed as a scalar field perturbation plus an arbitrary (weak) perturbation
corresponding to geodesically moving matter on the background. The key ingredient
in obtaining this system has been the appropriate selection of the 
gauge fixing condition given by (\ref{covgaugecond}) and (\ref{Bmu}).
The residual gauge invariance remaining under this  gauge fixing condition
has been imposed is given by the vector fields $\xi^{\mu}$ fulfilling the equation

\begin{equation}
\Box \xi_{\mu} + \,  2\, \Dot{\Omega} \, \partial_{\eta} \xi_{\mu}
+ \,  2 \Ddot{\Omega} \, \left( \xi_{\mu} - \, \delta_{\mu}^0 \, \xi_{0}   \right)
= \, 0  ~~,
\label{resgauge}
\end{equation}

that can be decomposed in the two simple equations

\begin{equation}
\left[ \Box \, + \, {\Dot{\Omega}}^2 - \, \Ddot{\Omega} \right]
 \left( a \, \xi^0 \right) =\, 0  ~~, 
\label{resgauget}
\end{equation}

and

\vspace{-0.2cm}
\begin{equation}
\left[ \Box \, + \, {\Dot{\Omega}}^2 + \, \Ddot{\Omega} \right]
\left( a \, \xi^i \right) =\, 0  ~~, 
\label{resgauges}
\end{equation}

\noindent
involving the same differential operators (and hence the same mode solutions) as
(\ref{cojoeqts}) and (\ref{cojoeqss}).
 
In the case in which the {\em free} perturbation $\delta T_{\mu}^{(F) \nu}$ 
vanishes, the gauge fixing condition plus the residual gauge invariance
reduce the number of physical degrees of freedom contained in the metric 
perturbation $\psi_{\mu}^{\hphantom{\mu} \nu}$ plus the scalar field perturbation 
$\delta \phi$, down two three:
a scalar perturbation plus two polarizations for gravitational waves on 
the background.

\vspace{0.75cm} 

{\bf 5. Universes with a linear equation of state}
\vskip 0.15cm

To obtain a definite expression for the retarded  Green's functions solving
equations  (\ref{cojoeqss}),  (\ref{cojoeqts}) and (\ref{cojoeqtt}), a
particular background model has to be chosen. In the case of universes
obeying a linear equation of state $p = \alpha \rho$, these equations take a 
particular simple form, and in addition the operators on the left hand
side of equations (\ref{cojoeqts}) and (\ref{cojoeqtt}) coincide.
Thus, in this case only two Green's functions for the two differential
operators 

\begin{equation}
\Box \, + \, {\Dot{\Omega}}^2 + \, \Ddot{\Omega} \,  \equiv \,
\Box + \, \frac{2 - 6\alpha}{(1+3\alpha)^2} \, \frac{1}{\eta^2} ~~,
 \label{opuno}
\end{equation}

\noindent
and 

\begin{equation}
\Box \, + \, {\Dot{\Omega}}^2 - \, \Ddot{\Omega} \, \equiv \,
\Box + \, \frac{6+6\alpha}{(1+3\alpha)^2} \, \frac{1}{\eta^2} ~~,
\label{opdos}
\end{equation}

\noindent
are required.

The Green's functions for these operators can be expressed as
superposition of homogeneous mode solutions, which are given by 
Bessel functions of indices $\nu_1 = \pm (3-3\alpha)/(2+6\alpha)$
and $\nu_2 = \pm (3\alpha + 5)/(2+6\alpha)$. Moreover in the particular 
cases of de Sitter ($\alpha = -1$) \cite{deSitter}, 
dust ($\alpha = 0)$ \cite{dust},  and 
radiation ($\alpha = 1/3$) dominated cosmological models,
the Bessel indices are the simplest half integers 
($\nu_1 = \pm 3/2, \nu_2 = \pm 1/2$),  
($\nu_1 = \pm 3/2, \nu_2 = \pm 5/2$), and
($\nu_1 = \pm 1/2, \nu_2 = \pm 3/2$)  respectively. 

Thus, to integrate the perturbations equations for de Sitter, dust, and radiation
backgrounds we only require the three retarded Green's functions solving

\begin{equation}
\left( \Box \, + \, \frac{A}{\eta^2} \right) \,
G_R^{(A)}(x,x') = - \delta^{(4)}(x-x') ~~,
\label{Greeneq}
\end{equation}

for $A = 0, 2, 6$ and where $x = (\eta, \vec{x})$. 

Then $G_R^{(0)}(x,x')$ is the well known retarded Green's function for
Minkowski, and  $G_R^{(2)}(x,x')$ and $G_R^{(6)}(x,x')$ can be obtained by 
elementary QFT methods as explained in \cite{deSitter}, with the result

\begin{eqnarray}
G_R^{(2)}(x,x') &  =  &
\frac{1}{4\pi\, |\vec{x} - \vec{x'}|} \, 
\delta (\eta -\eta' - |\vec{x} - \vec{x'}|) \nonumber \\
& & + \, 
\frac{1}{4\pi \, \eta \eta'}  \, \theta   (\eta -\eta' - |\vec{x} - \vec{x'}|) ~~, 
\label{green2}
\end{eqnarray}

\noindent
and

\begin{eqnarray}
G_R^{(6)}(x,x') &  = & \frac{1}{4\pi\, |\vec{x} - \vec{x'}|} \, 
\delta (\eta -\eta' - |\vec{x} - \vec{x'}|) \nonumber \\
& &  + \, 
\frac{3}{8\pi} \, \left[ \frac{1}{\eta^2} + \frac{1}{{\eta'}^2} 
- \frac{|\vec{x} - \vec{x'}|^2}{\eta^2 {\eta'}^2 } \right] \nonumber \\
& & \times  \theta (\eta -\eta' - |\vec{x} - \vec{x'}|)   ~~.
\label{green6}
\end{eqnarray}

Thus, $G_R^{(2)}(x,x')$ and $G_R^{(6)}(x,x')$ are given by the Minkowski
retarded Green's function with support on the past light cone plus 
an additional term with support in the interior of the past light cone.
Then the metric perturbation $\psi_{\mu}^{\hphantom{\mu} \nu}$ 
can be expressed as retarded potentials by means of these Green's functions.

\vspace{0.75 cm}
{\bf 6. Concluding renarks}
\vskip 0.15cm

We have analyzed cosmological perturbations on a spatially flat FRW background
given by a scalar field perturbation plus an arbitrary (weak) perturbation 
corresponding to test matter moving geodesically on this background.
We have shown that by an apropriate choice of the gauge the cosmological
perturbations obey decoupled hyperbolic equations. The class of perturbations 
considered have a high degree of generality, although it is not fully general
since a covariantly conserved tensor, linearly dependent in the metric 
perturbation, could still be added to the perturbing energy-momentum tensor.

On the other hand since the treatment and the gauge fixing condition 
that we have introduced is covariant with respect to finite coordinate
transformations on the background, it provides a very suitable framework
to develop linear quantum gravity around a spatially flat FRW background.

It is important to fully elucidate the relationships of the present
treatment with the standard Lifshitz-Bardeen treatment in terms of
gauge invariant perturbations and the decomposition of the metric perturbation
in scalar, vector and tensor parts.  This task together with more details 
of calulations and applications are being worked out and will be the
subject of a forthcomming paper. 



\vspace{0.75 cm}

{\bf  7. Acknowledgements}
\vskip 0.15cm
S. Kopeikin thanks the Departamento de F\'{\i}sica Te\'orica I, Universidad
Complutense de Madrid, for hospitality. J. Ram\'{\i}rez thanks the Department
of Physics and Astronomy,  University of Missouri-Columbia for hospitality.
We are grateful to  B. Mashhoon and A. L\'opez Maroto for valuable
discussions as well as A. Corman for helpful conversations.



\begin{thebibliography}{99}
\bibitem{isom} 
Since we are dealing with a flat FRW background, the six dimensional
isometry group of the background is generated by  three independent 
rotations, and three independent translations which are profitable
to do Fourier analysis of the perturbations.
\bibitem{conventions}
We use the same conventions for the signs of the metric and 
curvature tensors as in Weinberg's book \cite{Weinberg}.
\end{thebibliography}
\end{document}